\documentclass[%
apsrev,
pra,
 preprint,%
]{revtex4-1}
\usepackage{graphicx}
\usepackage{dcolumn}
\usepackage{bm}
\usepackage{physics} 
\usepackage[utf8]{inputenc}
\usepackage[T1]{fontenc}
\usepackage{mathptmx}
\usepackage{etoolbox}
\usepackage{xcolor}
\usepackage{amsmath}
\usepackage{amssymb}
\usepackage{caption}
\usepackage{subcaption}
\usepackage{float}
\usepackage{tikz}

\newcommand{\dbr}{\text{d}\mathbf{r}}

\newcommand{\br}{\mathbf{r}}
\newcommand{\dm}{\rho_{1}}
\newcommand{\dmrrp}{\dm(\br, \br')}
\newcommand{\dmrrpa}{\dm^{\alpha}(\br, \br')}
\newcommand{\dmrrpb}{\dm^{\beta}(\br, \br')}
\newcommand{\ddm}{\rho_{2}}
\newcommand{\gfrrp}{G(\mathbf{r}, \mathbf{r}'; \beta)}
\newcommand{\gfrrps}{G^{\sigma}(\mathbf{r}, \mathbf{r}'; \beta)}
\newcommand{\obeight}{\frac{1}{8}}

\newcommand{\obtwo}{\frac{1}{2}}
\newcommand{\twobthree}{\frac{2}{3}}

\newcommand{\ham}{\hat{\mathcal{H}}}
\newcommand{\vweiz}{ \frac{\nabla \rho(\br) . \nabla \rho(\br)}{\rho(\br)}}
\newcommand{\ef}{\epsilon_F} 
\newcommand{\etal}{\textit{et. al.}}

\newcommand{\cf}{\textit{cf.}}

\newcommand{\redsolidline}{\raisebox{2pt}{\tikz{\draw[-,red,solid,line width = 1.5pt](0,0) -- (5mm,0);}}}
\newcommand{\reddashedline}{\raisebox{2pt}{\tikz{\draw[-,red,dashed,line width = 1.5pt](0,0) -- (5mm,0);}}}
\newcommand{\bluesolidline}{\raisebox{2pt}{\tikz{\draw[-,blue,solid,line width = 1.5pt](0,0) -- (5mm,0);}}}
\newcommand{\bluedashedline}{\raisebox{2pt}{\tikz{\draw[-,blue,dashed,line width = 1.5pt](0,0) -- (5mm,0);}}}
\newcommand{\blacksolidline}{\raisebox{2pt}{\tikz{\draw[-,black,solid,line width = 1.5pt](0,0) -- (5mm,0);}}}
\newcommand{\blackdashedline}{\raisebox{2pt}{\tikz{\draw[-,black,dashed,line width = 1.5pt](0,0) -- (5mm,0);}}}
\newcommand{\purplesolidline}{\raisebox{2pt}{\tikz{\draw[-,purple,solid,line width = 1.5pt](0,0) -- (5mm,0);}}}
\newcommand{\tealsolidline}{\raisebox{2pt}{\tikz{\draw[-,teal,solid,line width = 1.5pt](0,0) -- (5mm,0);}}}

\begin{document}

\title{Kinetic energy density for open-shell systems: Analysis and development of a novel technique}
\author{Priya}
\affiliation{Department of Chemistry, Indian Institute of Technology Kanpur, Uttar Pradesh, India}
\author{Mainak Sadhukhan}
\email{mainaks@iitk.ac.in}
\affiliation{Department of Chemistry, Indian Institute of Technology Kanpur, Uttar Pradesh, India}

\begin{abstract}
 The quest for an approximate yet accurate kinetic energy density functional is central to the development of orbital-free density functional theory. While a recipe for closed-shell systems has been proposed earlier, we have shown that it cannot be na\"ively extended to open-shell atoms. In this present work, we investigated the efficacy of an \textit{ad-hoc} recipe to compute the kinetic energy densities for open-shell atoms by extending the methodology used for closed-shell systems. We have also analyzed the spin-dependent features of Pauli potentials derived from two previously devised enhancement factors. Further, we have proposed an alternate but exact methodology to systematically compute the kinetic energy density for atoms of arbitrary spin multiplicity.

\end{abstract}
\maketitle
\raggedbottom
\section{Introduction}

Kinetic energy density (KED) functional is the toughest bottleneck for developing an effective orbital-free density functional theoretic framework for atoms and molecules. 
However, most of the developments in that direction have not considered the effects of spin-polarization explicitly. In this work, we have extended some previously developed forms of kinetic energy density to open-shell systems and introduced a novel and exact route to derive kinetic energy density in a systematic manner.  

\par  The fundamental promise of density functional theory is that the total energy of an electronic ground state can be completely characterized by its one particle reduced density \cite{hohenKohn1964, Levy1979}. Despite this assurance, the analytical form of KED and inter-electronic repulsion potential as functionals of electron density is unknown. To bypass this problem, Kohn and Sham \cite{kohnsham1965} proposed a scheme where the density is partitioned into orbitals and computed from the approximated inter-electronic repulsion and kinetic energy density. The problem of kinetic energy, though hidden in KS scheme, remains an open challenge to this day\cite{WesolowskiWang2013}. 
 While in principle exact, Kohn-Sham equations cannot be applied to large systems due to prohibitive computational cost\footnote{For most of these methods diagonalization of extremely large matrices creates a bottle-neck. The diagonalization procedure scales as $N^3$ for an $N$-dimensional matrix}. In OFDFT, the total electron density can be computed from an Euler equation \cite{ParrYang1994, WesolowskiWang2013, Finzel2020} or imaginary time-evolution of a time-dependent hydrodynamical equation\cite{Deb1989, Roy1999}. As a result, the computational cost does not depend upon the number of electrons in the system explicitly, making the search for an accurate OFDFT method important. Recent efforts to employ OFDFT methods with the  machine learning algorithms \cite{mayeretal2020,ryczko2022} and pseudopotentials \cite{Xu2022} method manifest the usefulness of an accurate and general form of KED for simulations of large scale materials.

Several attempts have been formulated over the years to obtain an accurate description of exact kinetic energy density\cite{WesolowskiWang2013}. For solid-state systems, the formulation of adequately accurate KED has been achieved\cite{wang1999orbital,Constantin2019}. However, an accurate KED for atomic and molecular systems which produces atomic shell structures via self-consistent field calculation is lacking\cite{Finzel2020, Karasiev2015a}. 
  Almost all approximations of KED are based on the Thomas-Fermi model \cite{thomas1927calculation, fermi1927statistical}. For solid state systems generalized-gradient approximations (GGA)\cite{Constantin2018,Xia2015,Seino2018}, meta-GGA \cite{Constantin2018,cancio2017}, conjoint gradient correction approach\cite{Francisco2021}, as well as non-local response-based approaches \cite{ludena2018,salazar2016} enjoyed some success. Two points weighted density approximation of KED \cite{chakraborty2017two} and bi-functional-based methods are also being developed at present\cite{finzelmolecule, finzel2021molecule}. The other related quantity essential to self-consistently compute correct electron density is Pauli potential \cite{march1986local,levy1988exact}. This quantity has long been associated with proper electronic shell structure\cite{debghosh1983, Karasiev2015a, carter_2018}. Recently, shell-structure-based functionals are also showing promises\cite{finzel2015shell,fabiano2022kinetic}. 
  
 We arranged the article as follows. First, we have introduced (1) an \textit{ad-hoc} procedure, (2) a GGA-functional based procedure and (3) an exact methodology to compute the enhancement factors for spin polarized systems in section \ref{theory}. In section \ref{results} we have rigorously re-parameterized and analyzed the Pauli potentials derived from GGA functionals for open and closed-shell systems.  The paper concludes with section \ref{conclude}. 
 
 \section{Theory}\label{theory}
 The kinetic energy\footnote{the spinless density matrices is used throughout the present work} of an electronic system is given by 
 \begin{equation}\label{tsdef}
     T[\rho] = \int t(\br, \rho(\br)) \dbr
 \end{equation}
 where the KED is defined by (atomic units have been used throughout the article) 
\begin{equation}
    t(\br, \rho(\br)) =-\frac{1}{2}\nabla^2 \dmrrp\vert_{\br=\br'}
\end{equation}
For a single determinantal wave function the one particle density matrix can be written as 
\begin{equation}\label{dmab}
    \dmrrp = \dmrrpa + \dmrrpb
\end{equation}
where 
\begin{equation}\label{dmsumphi}
    \dmrrpa = \sum_{i\in \{\alpha\}}^{\epsilon_\alpha} (\phi^{\alpha}_i(\br'))^*\phi^{\alpha}_i(\br).
\end{equation}
Here $\epsilon_{\alpha}$ and $\phi^{\alpha}_i$ are highest occupied level of $\alpha$ spin orbitals and $i^{\text{th}}$ orbitals of $\alpha$ spin manifold, respectively. If we use Eq.\eqref{dmsumphi} for density matrix, where the orbitals are Kohn-Sham orbitals, then we get the corresponding kinetic energy $T_{s}[\rho]$.   

 In orbital-free density functional theory (OFDFT), the total kinetic energy functional is written as, 
 \begin{equation}\label{totalke_breakdown}
     T[\rho]= T_{vW}[\rho]+T_{corr}[\rho].
 \end{equation}
 Here, the first term 
 \begin{equation}
     T_{\text{vW}}[\rho] = \obeight \int \vweiz \dbr  
 \end{equation}
 is the von Weizs\"acker kinetic energy. The second term  
 \begin{equation}\label{tcorr}
     T_{corr}[\rho]= C_{TF}\int F(\br) {\rho(\br)}^{5/3} \dbr 
 \end{equation}
is known as modified Thomas-Fermi kinetic energy. For $F(\br) = 1 \forall \br$, $T_{corr}[\rho]=T_{TF}[\rho]$. $T_{TF}[\rho]$ is the bare Thomas-Fermi kinetic energy.  If we assume $T[\rho] =T_{s}[\rho]$ then $T_{corr}[\rho]$ is called the Pauli kinetic energy. The modulating function $F(\br)$ is also known as enhancement factor if we consider the Kohn-Sham limit of kinetic energy. Deb \etal \cite{Roy1999} proposed a parametric model of $F(\br)$ for closed shell atoms as a sum of a few Gaussian functions. Following  several compelling evidences\cite{ParrYang1994, Alonso1978}, they computed $F(\br)$
from Hartree-Fock kinetic energy density
\begin{widetext}
\begin{equation}\label{fr_debghosh}
    t_{\text{HF}}(\mathbf{r}, \rho) = -\frac{1}{4}\nabla^{2}\rho
+\frac{1}{8} \frac{\nabla \rho(\br) . \nabla \rho{(\br)})}{\rho(\br)} + C_{TF} F(\br) \rho^{5/3}(\br)
\end{equation}
\end{widetext}
The Hartree-Fock KED $t_{\text{HF}}$ is computed using the Hartree-Fock orbitals in Eq.\eqref{dmsumphi}.

Note that, the Kohn-Sham limit of kinetic energy is not the true kinetic energy and a part of it is dumped in exchange-correlation functional. In terms of Levy-Lieb constrained-search formulation of DFT, we can construct the Kohn-Sham equation from defining the extremum condition for $T_{s}$. This approach, however, gives rise to Wang paradox\cite{WesolowskiWang2013}. Nevertheless, in this article, we will use Pauli potential as a shorthand for $T_{corr}$. 

For self-consistent calculation of electron density, one can use the imaginary-time evolution of one-particle Deb-Chattaraj (DC) equation\cite{Deb1989}
\begin{widetext}
\begin{equation}\label{dceqn}
    \left(-\obtwo \nabla^2 + V_{Coul}(\br) + V_{XC}(\br) + \frac{5}{3}C_{TF}g(\br) \rho^{2/3}(\br)\right)\psi(\br,t) = i\frac{\partial \psi(\br,t)}{\partial t}.
\end{equation}
\end{widetext}
Here the hydrodynamical function $\psi(\br,t)$yields the electron density as $\rho(\br) = \abs{\psi(\br)}^2$. $V_{Coul}(\br)$ and $V_{XC}(\br)$ are electrostatic (of classical mechanical origin) and exchange-correlation (of quantum mechanical origin) potentials, respectively.  The last term in Eq.\eqref{dceqn} is known as the Pauli potential 
\begin{equation}\label{vp}
    V_{p}(\br) = \frac{\delta T_{corr}[\rho]}{\delta \rho(\br)}=\frac{5}{3} C_{TF} \rho^{2/3}(\br) g(\br),
\end{equation}
where $C_{TF}= \frac{3}{10} {(3\pi^{2})}^{2/3}$. In their works, Deb \etal. modeled $g(\br)$ as a sum of Gaussian functions. They have used one Gaussian function per shell. The same technique has recently been used by Finzel \cite{Finzel2020} to model the Pauli potential where the Gaussian functions were replaced by Slater-type functions. Recently efforts have been made to approximate the Pauli potential via a Pad\'e-type functional of $r$ as well as $\rho(\br) $\cite{finzel2015shell}. 

Eq.\eqref{totalke_breakdown} and subsequent derivation of Eq.\eqref{fr_debghosh}, however, assumes that the two-particle density matrix can be written as
\begin{equation}\label{d2_to_d1}
    \ddm(\br, \br') = \frac{1}{2}\left[\rho(\br)\rho(\br')-\dmrrp \dm(\br', \br)\right]
\end{equation}
and 
\begin{equation}\label{dm_equal}
    \dmrrp = \dm(\br', \br).
\end{equation}
The formulation of $F(\br)$ is then followed by defining the correlation factor 
\begin{equation}\label{cfunc}
      C(\br, \br')= \frac{2 \ddm(\br, \br')}{\rho(\br)\rho(\br')}-1.
\end{equation}
such that one can compute one-particle density matrix as 
\begin{equation}
    \dmrrp = \left(\rho(\br)\rho(\br')\right)^{1/2} \left(-C(\br, \br')\right)^{1/2}
\end{equation}

For homogeneous electron gas, the correlation factor is exactly known. A modification due to inhomogeneous electron density results a modified corrlation function, which in turn yields $F(\br)$. However, Eq.\eqref{dm_equal} and Eq.\eqref{d2_to_d1} are strictly correct for closed-shell systems. For an open shell system the two-particle density matrix for a single determinantal wave function can be written as \cite{ParrYang1994}
\begin{widetext}
\begin{equation}\label{dab2_to_dab1}
     \ddm(\br, \br') = \frac{1}{2}\left[\rho(\br)\rho(\br')-\dm^{\alpha\alpha}(\br, \br') \dm^{\alpha\alpha}(\br', \br)-\dm^{\beta\beta}(\br, \br') \dm^{\beta\beta}(\br', \br)\right].
\end{equation}
\end{widetext}
 Eq.\eqref{dab2_to_dab1}, unlike Eq.\eqref{d2_to_d1}, cannot be inverted to obtain the expressions for $\dm^{\alpha}(\br, \br')$ or $\dm^{\beta}(\br, \br')$ separately. Moreover, since 
 \begin{equation}
 \dmrrp \neq \sqrt{(\dm^{\alpha\alpha}(\br, \br'))^2+(\dm^{\beta\beta}(\br, \br'))^2}
 \end{equation}
 due to Eq.\eqref{dmab}, we cannot use Eq.\eqref{cfunc} to define the correlation function for open-shell systems. Due to the same reasons, a straightforward extension of spin-polarized Pauli potential is also unavailable. 
 
 \subsection{An \textit{ad-hoc} definition of spin-polarized enhancement factors}\label{adhoc}
  Extensions of Eq.\eqref{tcorr} and Eq.\eqref{vp} for open-shell systems, are therefore, necessary for a generalized description of atoms and molecules. The first attempt, albeit an \textit{ad-hoc} one, is inspired by the spin-polarized extension of von Weizs\"acker and Thomas-Fermi kinetic energy\cite{ParrYang1994} given by
  \begin{equation}\label{kealphabeta}
      T_{\text{vW}}[\rho^{\alpha}, \rho^{\beta}] = \frac{1}{8}  \int \frac {|{ \nabla \rho ^{\alpha}(\br)}|^{2}}{\rho^{\alpha}(\br)} d\mathbf{r} +\frac{1}{8} \int \frac {|{ \nabla \rho ^{\beta}(\br)}|^{2}}{\rho^{\beta}(\br)} d\mathbf{r}
  \end{equation}
and 
 \begin{equation}
     T_{TF}[\rho^{\alpha}, \rho^{\beta}] = 2^{\twobthree} C_{TF} \int {(\rho^{\alpha}(\br))}^{5/3} +{(\rho^{\beta}(\br))}^{5/3} \hspace{0.1cm} d\mathbf{r},
 \end{equation}
  respectively. We therefore define spin-polarized Pauli kinetic energy
  \begin{eqnarray}
     T_{corr}[\rho^{\alpha}, \rho^{\beta}]&=& \sum_{\sigma} T_{corr}^{\sigma}[\rho^{\sigma}]\\
     T_{corr}^{\sigma}[\rho^{\sigma}] &=& C_{TF} \int F^{\sigma}(\br){(\rho^{\sigma}(\br))}^{5/3}
  \end{eqnarray}
  where spin variable $\sigma \in \{\alpha, \beta\}$. Note that the factor $2^{2/3}$ is absorbed inside $F^{\sigma}(\br)$. The spin-polarized Pauli potential is therefore can be computed by 
  \begin{equation}
        V_{p}^{\sigma}(\br) = \frac{\delta T_{corr}^{\sigma}[\rho]}{\delta \rho^{\sigma}(\br)}=\frac{5}{3} C_{TF} (\rho^{\sigma}(\br))^{2/3} g^{\sigma}(\br),
  \end{equation}
  
  \subsection{Generalized gradient approximation-based Pauli potentials} \label{gga_paulipot}
  Different classes of enhancement factors \cite{Francisco2021} have been introduced over the years. In a recent work \cite{Constantin2019} some of the enhancement factors based on generalized gradient approximation (GGA) have been analyzed for semiconductors and metals. Here we will consider two of their simplest enhancement factors and analyze the corresponding Pauli potentials for open and closed-shell systems. These two enhancement factors are given by 
  \begin{enumerate}
      \item the linear function
      \begin{equation} \label{linef}
    F_{L}(s;\lambda) = 1- \frac{5}{3} s^{2}(\lambda -1)
      \end{equation}
     and 
     \item the Pauli-Gaussian(PG) enhancement function
     \begin{equation} \label{gausef}
       F_{E}(s;\mu) = e^{ -\mu s^{2}}
     \end{equation}
  \end{enumerate}
 where $\lambda$ and $\mu$ are free parameters. Both of them have been expressed in terms of scaled gradient of density 
 \begin{equation}
     s= \frac{\abs{\nabla \rho(\br)}}{C_{0} \rho^{4/3}(\br)} 
 \end{equation}
  where $C_{0}=2(3 \pi^{2})^{1/3}$. Other forms of $F(\br)$ and further modifications by introducing scaled Laplacian
\begin{equation}
  q= \frac{\nabla^{2} \rho}{[4 (3 \pi^{2})^{2/3} \rho^{5/3}]}
\end{equation}
have also been considered recently\cite{Brack1976, cancio2016,cancio2017}. However, we will not consider them in this work.

The Pauli potential and therefore the corresponding $g(\br)$ can be computed by
\begin{widetext}
\begin{equation}
    \frac{1}{C_{TF}}\frac{\delta T_{corr}[\rho]}{\delta \rho(\br)}= \frac{\partial(F(\br) \rho(\br)^{5/3})}{\partial \rho} - \nabla \left(\frac{\partial(F(\br) \rho(\br)^{5/3})}{\partial (\nabla \rho)} \right)
+\nabla^{2} \left( \frac{\partial(F(\br)\rho(\br)^{5/3}) }{\partial (\nabla^{2} \rho)}  \right) - \hdots 
\end{equation}
\end{widetext}

 The Pauli potentials corresponding to Eq.\eqref{linef} and Eq.\eqref{gausef} are
 \begin{widetext}
 \begin{eqnarray}
     g_{L}(\br)= \left(\frac{3}{5}F_{L}(\br)+\frac{2}{5}+ \frac{6 \mu}{{C_{0}}^{2}} \frac{\nabla^{2} \rho(\br)}{\rho^{5/3}(\br)} \right) \label{gl}\\
     g_{E}(\br) = F_{E}(\br)\left(\frac{2\mu p}{5}+\frac{16 \mu^2 p^2}{5}+ \frac{6 \mu(1-2\mu p)}{5{C_{0}}^{2}} \frac{\nabla^{2} \rho(\br)}{\rho^{5/3}(\br)}+1 \right)\label{ge} 
 \end{eqnarray}
 \end{widetext}
 where $p=s^2$. The $\mu$ in Eq.\eqref{gl} is defined as $\mu = \frac{5}{3}(\lambda-1)$. Note that $s$ is a dimensionless quantity. These expressions can therefore be extended to spin-polarized cases phenomenologically by replacing $\rho(\br)$ by $\rho^{\sigma}(\br)$ and  $s$ by $s_{\sigma} 2^{-1/3}$.  Here 
 \begin{equation}
     s_{\sigma} = \frac{\abs{\nabla \rho^{\sigma}(\br)}}{C_{0} (\rho^{\sigma}(\br))^{4/3}} 
 \end{equation}
 
 \subsection{An exact and systematic method using Green's function}
 Now we will lay an exact road map for deriving the enhancement factor and therefore Pauli potential using Green's function (GF) method. We start from defining the Green's function $\gfrrp$ as the Laplace transform of the density matrix \cite{ParrYang1994}
 \begin{equation}
    \gfrrp = \int_{0}^{\infty} \text{d}\ef e^{-\beta \ef} \dmrrp.
 \end{equation}
 Since the Laplace transform is a linear transform, we can extend the definition of Green's function to spin-polarized cases in a straightforward manner (\cf Eq.\eqref{dmab}). We can obtain density matrix for a given spin multiplicity via Bromwich integral
 \begin{equation}\label{bromwich}
  \dm^{\sigma}(\br,\br') = \lim_{T\to \infty}\frac{1}{2\pi i} \int_{\gamma-iT}^{\gamma+iT}\frac{\text{d}\beta}{\beta}e^{\beta \ef} \gfrrps
 \end{equation}
 where $\gamma \in \mathbb{R}_{>} $. As a result, $\gfrrps$ can be written as  
 \begin{widetext}
\begin{equation}
    \gfrrps = \bra{\br}e^{-\beta \ham}\ket{\br'}\vert_{\sigma} = \sum_{i\in \{\sigma\}}^{\infty}(\phi^{\sigma}_i(\br'))^*\phi^{\sigma}_i(\br) e^{-\beta \epsilon_i^{\sigma}}. 
 \end{equation}
 \end{widetext}
 where $\ham$ is the corresponding Hamiltonian with eigenvalues $\{\epsilon_{i}^{\sigma}\}$ and eigenfunctions $\{\phi_{i}^{\sigma}\}$. We will drop the superscript $\sigma$ for Green's functions in this section for brevity. However, every Green's function discussed onward will be assumed as spin-polarized.    
 
 We can write the Hamiltonian of an atom as 
 \begin{equation}
     \ham = \underbrace{\ham_0+\ham_{Z}}_{\ham_{H}} + \hat W
 \end{equation}
where $\ham_0$, $\ham_{Z}$ and $\hat W$ are free-particle Hamiltonian, electron nucleus attraction potential and inter-electronic repulsion potentials respectively. $\ham_{H} = \ham_0+\ham_{Z}$ is the sum of Hamiltonians of a  Hydrogenic atom with $Z$. Consequently, the density matrix corresponding to $\ham_{H}$  produces von Weizs\"acker kinetic energy with correction which can be computed in a systematic manner \cite{sim2003}. We define two more Green's functions 
\begin{equation}
 G^{H}(\br, \br'; \beta) = \bra{\br}e^{-\beta \ham_{H}}\ket{\br'}
\end{equation}
and 
\begin{equation}
 G^{0}(\br, \br'; \beta) = \bra{\br}e^{-\beta \ham_{0}}\ket{\br'}
\end{equation}
 corresponding to the Hydrogenic system and free-particle systems respectively. Note that $\ham_{HEG}=\ham_0 + \hat W$ describes the homogeneous electron gas limit and we will call the corresponding Green's function $G^{HEG}(\br, \br';\beta)$. 
 
 A connection between these Green's functions can be obtained 
 \begin{equation}\label{g_rrp_rep}
    G(\br, \br'; \beta)- G^{H}(\br, \br'; \beta) = \bra{\br}\hat O\ket{\br'}
 \end{equation}
 where 
 \begin{equation}
     \hat O = e^{-\beta \ham} - e^{-\beta \ham_{H}}.
 \end{equation}
 Using Zassenhaus formula \cite{CasasNadinic2012, magnus1954} we can write
 \begin{equation}\label{hato_zassenhaus}
     \hat O = - e^{-\beta \ham_{H}}\left( 1-e^{-\beta \hat W} e^{-\frac{\beta^2}{2} \comm{\ham_H}{\hat W}} \mathcal{O}(\beta^3) \right).
 \end{equation}
We find
\begin{equation}
    \comm{\ham_H}{\hat W}=\comm{\ham_0}{\hat W} 
\end{equation}
since $\comm{\ham_Z}{\hat W}$ commutes. As a result, all commutator in Eq.\eqref{hato_zassenhaus} can be replaced by $\comm{\ham_0}{\hat W}$. As a result,
\begin{equation}\label{hatohydr_heg}
     \hat O = - e^{-\beta \ham_{H}} \left( 1-e^{-\beta \hat W} e^{-\frac{\beta^2}{2} \comm{\ham_0}{\hat W}} \mathcal{O}(\beta^3) \right) = e^{-\beta \ham_{H}} e^{\beta \ham_{0}}\left(e^{-\beta (\ham_0+\hat W)} - e^{-\beta \ham_{0}}\right) .
\end{equation}
Using Eq.\eqref{hatohydr_heg} in Eq.\eqref{g_rrp_rep} and employing resolution of identity, we obtain 
\begin{equation}\label{resoid}
    \gfrrp = G^{H}(\br, \br'; \beta)+ \int \dbr'' \bra{\br}e^{-\beta \ham_{H}}\ket{\br''}\bra{\br''}\sum_{n=0}^{\infty} \frac{\beta^n}{n!}\ham_{0}^{n}\left(e^{-\beta (\ham_0+\hat W)} - e^{-\beta \ham_{0}}\right) \ket{\br'}.
 \end{equation}
Eq.\eqref{resoid}, similar to a Dyson equation,  presents the total Green's function as an infinite series and is an exact result, leading to a systematic way to compute the total Green's function. Using the definition of Green's functions and considering only leading order term ($n=0$) of Eq.\eqref{resoid}, we obtain
\begin{equation}\label{leadingordergf}
    \gfrrp \approx G^{H}(\br, \br'; \beta)+ \int \dbr'' G^{H}(\br, \br'';\beta)\left(G^{HEG}(\br'', \br';\beta)-G^{0}(\br'', \br';\beta)\right).
\end{equation}
Since both $\ham_{H}$ and $\ham_{HEG}$ can be written as $\ham_0 + \text{other terms }$, we can again apply Zassenhauss formula for $G^{H}(\br, \br'; \beta)$ and $G^{HEG}(\br, \br'; \beta)$ to obtain Green's functions as
\begin{eqnarray}
    G^{H}(\br, \br'; \beta) = G^{0}(\br, \br'; \beta)  + \ldots\label{gh_g0}\\
    G^{HEG}(\br, \br'; \beta) = G^{0}(\br, \br'; \beta)  + \ldots\label{gheg_g0}
\end{eqnarray}
Using Eqs.\eqref{gh_g0} and \eqref{gheg_g0} in Eq.\eqref{leadingordergf}, we obtain
\begin{equation}
    \gfrrp \approx G^{H}(\br, \br'; \beta)+ \int \dbr''G^{0}(\br, \br''; \beta)G^{0}(\br'', \br'; \beta).
\end{equation}
Using the definition of Green's function used here we find therefore the leading order term of the Green's function is 
\begin{equation}\label{gtog0}
     \gfrrp \approx G^{H}(\br, \br'; \beta)+ G^{0}(\br, \br'; 2\beta). 
\end{equation}
It can be seen from Eq.\eqref{bromwich} that 
\begin{equation}\label{green0}
      \int_{\gamma-iT}^{\gamma+iT}\frac{\text{d}\beta}{\beta}e^{\beta \ef} G^{0}(\br, \br'; 2\beta) = \int_{\gamma-iT}^{\gamma+iT}\frac{\text{d}(2\beta)}{(2\beta)}e^{(2\beta) \ef}\sum_{k=0}^{\infty} \frac{(-\beta\ef)^k}{k!}G^{0}(\br, \br'; 2\beta)
\end{equation}
Again using the leading-order term of Eq.\eqref{green0} in Eq.\eqref{gtog0}, we find that 
\begin{equation}
    \dmrrp = \dm^{H}(\br, \br') + \dm^{0}(\br, \br') + \mathcal{O}(\beta)
\end{equation}
where $\dm^{H}(\br, \br')$ and $\dm^{0}(\br, \br')$ are density matrices for a Hydrogenic atom and a free-particle system, respectively. Clearly, the form of kinetic energy from this approach turns out to be 
\begin{equation}\label{t_eq_tvwplusttf}
    T[\rho] = T_{vW}[\rho] + T_{TF}[\rho]+ \text{higher order terms}.
\end{equation}
The higher order terms now can be considered systematically. However, such a task will require significant mathematical exercise and will be considered later. Eq.\eqref{t_eq_tvwplusttf} is the form described by Carter \etal\cite{wang1999orbital, carter_2018}.

 \section{Results and discussion}\label{results}
 We have considered two noble gas atoms (Neon, Argon) and four first-row elements (Lithium, Beryllium, Boron and Carbon) as our representative systems. Three of them (Ar, Ne, Be) are spin-paired while the other three (Li, B, C) are spin-polarized. Among them, Li also has zero beta spin in second shell. This variety of electron structure is necessary to understand the effects of spin-multiplicity on the Pauli potential. For larger atoms, more intricate effects such as spin-orbit coupling and relativistic corrections become important as well. These effects are outside of the scope of this work. We will consider them in future works.

 All densities are computed from parametrized Slater functions for the Hartree-Fock wave functions \cite{Clementi1974}. All benchmark results for $F(\br)$ and $g(\br)$ (dubbed $F_{HF}(\br)$ and $g_{HF}(\br)$, respectively onward) are computed using the parameters provided in Ref.\cite{Roy1999}.

 \subsection{Optimization of free parameters for $g_{E}(\br)$ and $g_{L}(\br)$}
 
 We begin by optimizing the value of parameter $\mu$ used in Eqs.\eqref{gl} and \eqref{ge}. To accomplish that, we have computed a metric of deviation from the benchmark $g_{HF}(\br)$ as
 \begin{equation}
     E(\mu) = \int_{0}^{\infty} r^2 \abs{g_{HF}(\br)-g_{model}(\br)}\text{d}r.
 \end{equation}
  A value of $\mu$ which minimizes $E(\mu)$ can be chosen as the optimum choice. We have chosen Ne and Ar atoms as our benchmark systems. Accurate $g_{HF}(\br)$ for them are available as a linear combination of parametrized Gaussian functions\cite{Roy1999, debghosh1983}. These parameters were set to reproduce the Hartree-Fock level electron density upon self-consistent field calculations.
  
\begin{figure}[h]
      \centering
      \includegraphics[scale=0.6, angle=270]{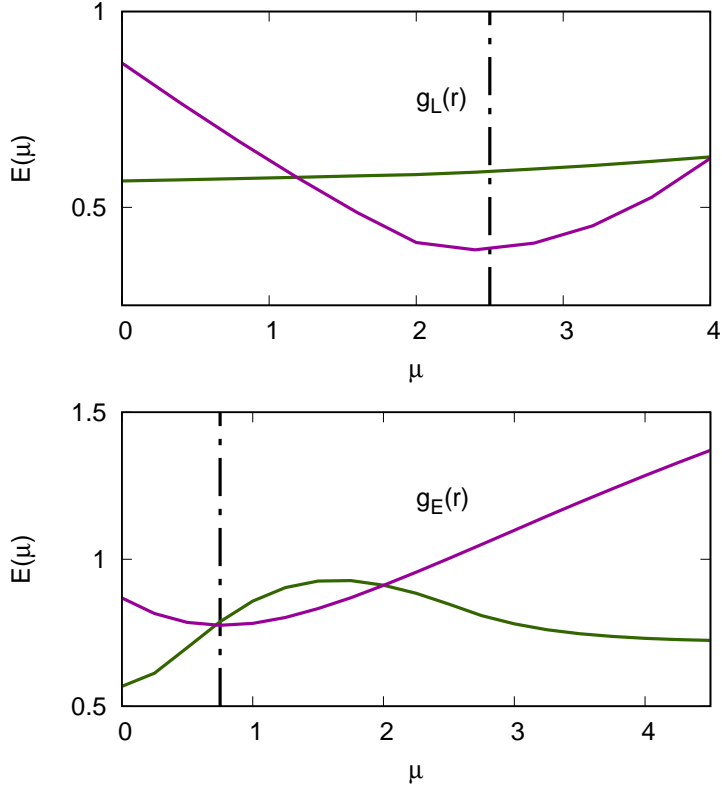}
     \caption{Variation of $E(\mu)$ with free parameter $\mu$ for $g_{L}(\br)$ (top panel) and $g_{E}(\br)$ (bottom panel). $E(\mu)$ for Ar (purple solid line \protect\purplesolidline) shows more prominent minimum than that of Ne (teal solid line \protect\tealsolidline) atom. Black dashed vertical line indicates the chosen value for $\mu$ for both $g_E$ and $g_L$. } \label{fig:Emu_vs_mu}
  \end{figure}

  Variations of $E(\mu)$ as a function of $\mu$ are presented in Fig.\ref{fig:Emu_vs_mu}. 
  
  For Ar atom, the minima are clear for both $g_{L}(\br)$ and $g_{E}(\br)$. For Ne, such clear minimum is not pronounced. Therefore, we have chosen the value for $\mu$ which minimizes $E(\mu)$ for Ar atom. It is clear that the parameter $\mu$ is actually system-dependent and therefore is not transferable. A better understanding of system dependence of $\mu$ is required if these enhancement factors are to be used for atomic systems. For all our subsequent results, $\mu=2.5$ for $g_{L}(\br)$ and $\mu=0.75$ for $g_{E}(\br)$ have been used irrespective of the systems. 
 \subsection{Comparison between different $g(\br)$ for closed-shell systems}\label{ArNe}
\begin{figure}[H]
 \centering
\includegraphics[angle=270, scale=0.7]{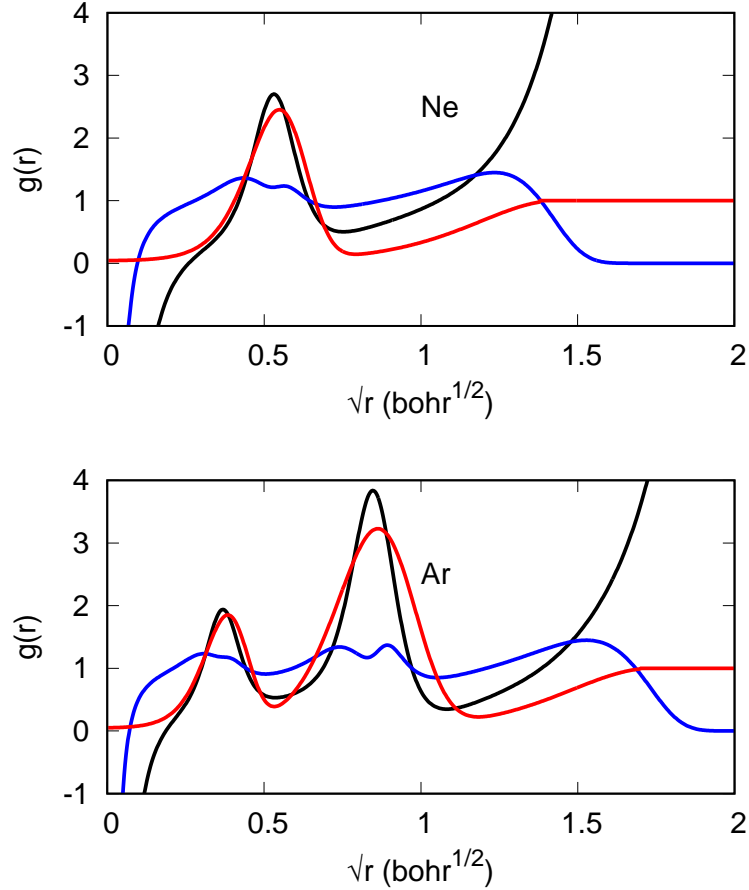}
 \caption{Comparison of $g_{HF}(\br)$ (red solid line \protect\redsolidline ) with  $g_{L}(\br)$ (black solid line \protect\blacksolidline) and $g_{E}(\br)$ (blue solid line \protect\bluesolidline). The top panel shows results for Neon and the bottom panel depicts results for Argon}\label{fig:g_comparison}
\end{figure}

 Having optimized the parameters, we now examine the features of each $g(\br)$ for closed shell atoms. We have compared them with the benchmark $g_{HF}(\br)$ functions mentioned above (Fig.\ref{fig:g_comparison}). In this plot both $g_{L}(\br)$ and $g_{E}(\br)$ do not behave well for $r \to 0$. A possible reason for this behavior is the Laplacian terms in Eq.\eqref{gl} and Eq.\eqref{ge}. 
$g_{L}(\br)$ (black solid line) mimics the shell structure for both Ne and Ar atoms more accurately. However, for large $r$, the function diverges quickly. Here large values of $r$ signify the distances where both the electron density and its gradients are very small. As a result, our computed $g(\br)$ is also susceptible to numerical inaccuracies. On the other hand, $g_{E}(\br)$ (blue solid line) shows very different profile while it is more well-behaved at large values of $r$. It appears that, although simpler in form, $g_{L}(\br)$ produces better shell-structures than the more complicated $g_{E}(\br)$
contrary to bulk systems\cite{Constantin2009, Constantin2019}.  
 
 \subsection{$g_{L}(\br)$ applied to open-shell systems }
\begin{figure}[H]
    \centering
    \includegraphics[scale=0.5, angle=270]{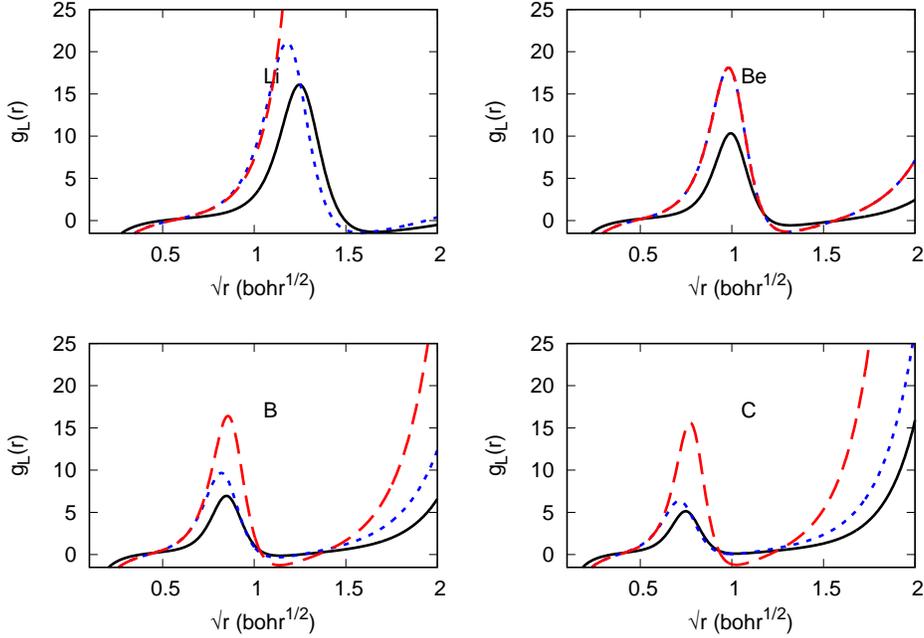}
    \caption{ Comparison of  $g_{L}(\br)$ for open shell atoms.Here   $g_{L}^{\alpha}(\br)$ (blue dashed line \protect\bluedashedline ) , $g_{L}^{\beta}(\br)$ (red dashed line \protect\reddashedline) and $g_{L}(\br)$ (black solid line \protect\blacksolidline). The top panel shows results for Lithium(first plot) and Beryllium(2nd plot), the bottom panel is for Boron and carbon.     }
    \label{fig:gl_comparision}
\end{figure}

 Following our analysis for closed-shell atoms in the previous section, we now examine $g_{L}(\br)$ for open-shell atoms. For such systems, we examined $g_{L}(\br)$, $g_{L}^{\alpha}(\br)$ and $g_{L}^{\beta}(\br)$ for the total density, $\alpha$ spin density and $\beta$ spin density, respectively (Fig.\ref{fig:gl_comparision}). For all four cases, $g_{L}^{\beta}(\br)$ showed the highest peak followed by $g_{L}^{\alpha}(\br)$ and $g_{L}(\br)$. This trend indicates that the peak height is inversely proportional to the number of electrons considered. This is also corroborated by Fig.\ref{fig:g_comparison} which clearly shows that Neon has the lowest peak height among all first-row elements considered. The peak height corresponding to the inner-shell of Argon atom is even lower supporting our observation. Also, the peak positions shift to smaller distances as the number of electrons increases, showing the contraction of electron density with increasing nuclear charge\cite{atomic radius}. As mentioned before  $g_{L}(\br)$ diverges for large value of $r$, explaining the diverging $g_{L}^{\beta}(\br)$ (no $\beta$ electron in $2s$) for Li.  For Be the $g_{L}^{\alpha} =g_{L}^{\beta}$, as expected. B and C have same number of $\beta$ electrons and hence exhibit very similar $g_{L}^{\beta}$. For C atom, $g^{\alpha}_{L}(\br)$ is very similar to that of $g_{L}(\br)$. While the shell structure is clearly visible for all four atoms, note that the $g_{L}$ values sometimes fall below zero, rendering the corresponding Pauli potential negative. This is a serious drawback of this $g_{L}(\br)$. 

\begin{figure}[h]
    \centering
    \includegraphics[scale=0.4, angle=270]{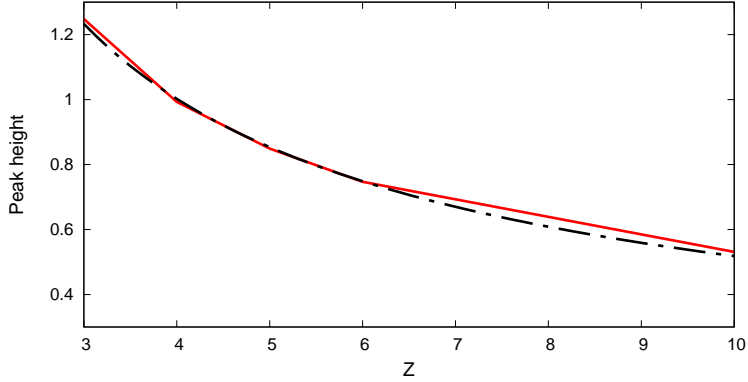}
    \caption{Variation of peak height with atomic number $(Z)$  for $g_{L}(\br)$ (red solid line \protect\redsolidline ) with fitted function (black dashed line \protect\blackdashedline)}
    \label{fig:peak_vs_z}
\end{figure}
Next we examine the variation of peak heights of $g_{L}(\br)$ with atomic number $Z$ (same as electron number $N$ here) for first-row elements (Fig.\ref{fig:peak_vs_z}). As expected, the peak height decreases with $Z$. To quantify this behaviour, we have fit the data with a monomial function $f(x)= ax^{b} $ where  $a=2.74$ and $b=-0.72$ indicating a $\approx Z^{-3/4}$ variation\cite{gazquez1982} \footnote{The root mean squared error is $0.01249$ and the reduced $\chi^2$ is $0.00015$.}.

 \subsection{$g_{E}(\br)$ applied to open-shell systems }
 
\begin{figure}[h]
    \centering
    \includegraphics[scale=0.5, angle=270]{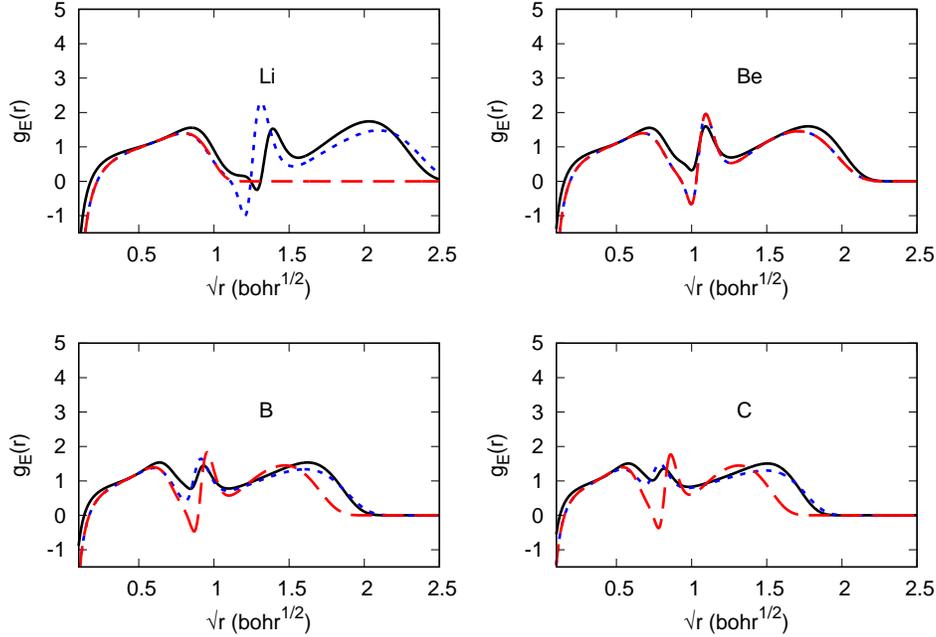}
    \caption{Comparison of  $g_{E}(\br)$ for open shell atoms.Here   $g_{E}^{\alpha}(\br)$ (blue dashed line \protect\bluedashedline ) , $g_{E}^{\beta}(\br)$ (red dashed line \protect\reddashedline) and $g_{E}(\br)$ (black solid line \protect\blacksolidline). The top panel shows results for Lithium(first plot) and Beryllium(2nd plot), the bottom panel is for Boron and carbon. }
    \label{fig:ge_comparison}
\end{figure}

 Next, we analyze the effects of $g_{E}(\br)$ for open-shell atoms (Fig.\ref{fig:ge_comparison}). Similar to the cases discussed in the previous section, we considered $g_{E}(\br)$, $g_{E}^{\alpha}(\br)$ and $g_{E}^{\beta}(\br)$ for the total density, $\alpha$ spin density and $\beta$ spin density, respectively. Following the same trend already observed in section\ref{ArNe}, we find that $g_{E}(\br)$ is well-behaved for large $r$. However, the absence of any clear shell structures, akin to $g_{L}(\br)$, obscures the applicability of Eq.\eqref{ge} in Pauli potential. Unlike the divergence observed for $g_{L}^{\beta}(\br)$ for Li, here we see that the $g_{E}^{\beta}(\br)$ settles to zero for $r \ge 1.2$. The problem of negative values for $g_{E}(\br)$ is more prominent here compared to $g_{L}(\br)$. The other features are shown by $g_{E}^{\alpha}$ and $g_{E}^{\beta}$ are not very clear at present. A more careful analysis of Eqs. \eqref{gl} and \eqref{ge} is therefore required to develop better Pauli potentials in the future.  
 
 \section{Conclusion}\label{conclude}
 
 In summary, we have (1) pointed out the mathematical difficulty to extend a previously devised derivation for enhancement factors for open-shell atoms, (2) analyzed Pauli potentials derived from two previously devised enhancement factors, and (3) developed an exact and novel method to compute the enhancement factors which do not depend on the spin-multiplicity of the system. To achieve this we employed Green's function technique to obtain a well-defined but infinite series expression for density matrix. We found that the Pauli potentials derived from previously used enhancement factors does not always meet the expected asymptotic behaviors. Also, the linear enhancement factor produces the atomic shell structure better compared to the more complicated Gaussian enhancement factors. However, both of these functions fail to provide a correct description of $g(\br)$ near the nucleus. Further studies are required to understand the reasons and remedies for this problem. Moreover, we have found that the shell structure in $g(\br)$ corresponding to linear enhancement factor is very sensitive to spin multiplicity.  
 This work raises some new questions as well. The possibility to derive adequately accurate yet analytically closed forms of enhancement factor and Pauli potential is the most significant among them. While the Green's function method is promising, the convergence of the infinite series remains an open question. Description of Pauli potential for molecular systems is another problem that may be taken up in the future. Furthermore, the effect of the bond-breaking and bond-making process on these quantities are yet to be explored. We will be addressing these questions in our future works.  
 \begin{acknowledgments}
 Priya acknowledges a prime minister's research fellowship for financial supports. 
 MS acknowledges IIT Kanpur initiation grant no. IITK/CHM/2018419 and SERB startup research grant no. SRG/2019/000369 for partial computational supports and Debashree Manna for numerous helpful discussions. 
 \end{acknowledgments}
 
 \section*{dedication} 
 This work is dedicated to Prof. B. M. Deb, one of earliest who dreamed up the success of density-based description of quantum chemistry. Also, one of us has found him as a friend, philosopher and guide for over a decade now.   

\bibliographystyle{apsrev4-1}
\bibliography{MP}
\end{document}